\documentclass[aps,prx,reprint,showpacs,showkeys,noeprint,longbibliography,superscriptaddress]{revtex4-2}
\usepackage{graphicx}
\usepackage{amsmath}
\usepackage{array}
\usepackage{amsfonts}
\usepackage{float}
\usepackage{xcolor}
 \usepackage{cmap}
 \usepackage[T1]{fontenc}
 \usepackage{textcomp}
 \usepackage{amsmath}
 \usepackage{latexsym}
 \usepackage{float}
  \usepackage{amssymb}
 \usepackage{graphicx}
 \usepackage{hyperref}
 \usepackage{xcolor}
 \usepackage{url}
 \usepackage{booktabs,tabularx,dcolumn}

\def\ba{\begin{eqnarray}}
\def\ea{\end{eqnarray}}
\def\be{\begin{equation}}
\def\ee{\end{equation}}
\def\bm{\begin{math}}
\def\me{\end{math}}

\newcommand{\dummy}

\makeatletter
\newcommand{\fmarki}{*}
\newcommand{\fmarkii}{\ensuremath{\dagger}}
\newcommand{\fmarkiii}{\ensuremath{\ddagger}}
\newcommand{\fmarkiv}{\ensuremath{\mathsection}}
\newcommand{\fmarkv}{\ensuremath{\mathparagraph}}
\newcommand{\fmarkvi}{\ensuremath{\|}}
\newcommand{\fmarkvii}{**}
\newcommand{\fmarkviii}{\ensuremath{\dagger\dagger}}
\newcommand{\fmarkix}{\ensuremath{\ddagger\ddagger}}
                
\def\@fnsymbol#1{{\ifcase#1\or \fmarki\or \fmarkii\or \fmarkiii\or \fmarkiv\or \fmarkv\or \fmarkvi\or \fmarkvii\or \fmarkviii\or \fmarkix \else\@ctrerr\fi}}
\makeatother

\begin{document}
%\title{Active or passive tracer coupled to active or passive bath}
% \title{Inertial Effects on the Spontaneous Velocity Alignment of Active Particles}

\title{Spontaneous Micro Flocking of Active Inertial Particles without Alignment Interaction }

%\title{Inertial Effects on Spontaneous Micro-flocking in Active Particles without Alignment Interaction }

\author{ Subhajit Paul} \email[]{spaul@physics.du.ac.in, subhajitphys@gmail.com}
\affiliation{International Center for Theoretical Sciences, Tata Institute of Fundamental Research, Bangalore 560089, India} \affiliation{Department of Physics and Astrophysics, University of Delhi, Delhi 110007, India }
\author{ Suman Majumder} \email[]{suman.jdv@gmail.com}
\affiliation{Amity Institute of Applied Sciences, Amity University Uttar Pradesh, Noida 201313,
	India}
\author{Wolfhard Janke}\email[]{wolfhard.janke@itp.uni-leipzig.de}
\affiliation{Institut  f\"{u}r Theoretische Physik, Universit\"{a}t Leipzig, IPF 231101, 04081 Leipzig, Germany}
\date{\today}

\begin{abstract}

%Observation of spontaneous velocity alignment or flocking has always been a challenge for spherical active Brownian particles. 

%In this work, for a system of underdamped active particles, we try to  break the myth that explicit alignment interaction is necessary to realize flocking transition in a system of spherical particles. 

Observing spontaneous velocity ordering or flocking during motility induced phase separation (MIPS) in a system of spherical active Brownian particles without alignment interaction is challenging. 
We take up this problem by performing simulations of spherical active inertial particles with purely repulsive potential in presence of thermal noise and absence of any explicit alignment interaction. Our results not only show the presence of MIPS,
but also reveal a micro-flocking transition. We characterize this transition in terms of a velocity order parameter as well as a characteristic length scale derived from the spatial correlation of the velocities.

\end{abstract}


\maketitle 
 Dynamics of many living objects, such as fishes, birds, bacteria,  covering a wide range of length scale are modeled as ``active'' particles \cite{chate_08,roman,marchetti_rmp13,elgeti_15,cates1,bechinger_16,shaeb_20,lowen_20,caprini_21,vicsek_95,gregoire2004,sokolov2007,chate_2008,mishra2,hagen_11,fily_12,bialke_12,redner_13,das_17,dhar19,mandal_19,paul1_21,bera_22}. For them dissipation and consumption of energy happen at the level of single particle. On top of that, existence of self propulsion makes both single-particle behavior and their collective motion entirely different from passive particles \cite{vicsek_95,gregoire2004,chate_2008,mishra2,hagen_11,fily_12,bialke_12,redner_13,das_17,mandal_19,paul1_21,dhar19}.  Models with these characteristics invoked, on introduction of interaction among the particles make it an appropriate candidate to explore various  nonequilibrium pattern formations and emergence of flocking  in the real biological systems \cite{marchetti_rmp13,elgeti_15,bechinger_16,fily_12,redner_13,mandal_19,paul1_21,cates1,bialke_12,das_17,sokolov2007}.  
 \par
 The minimal model of Vicsek \emph{et al.}\ was the first to explore such a collective phenomenon \cite{vicsek_95}. For sufficiently high density and low thermal noise, in presence of a local velocity-alignment interaction among the particles, the model shows a transition from a disordered to an ordered state where particles move coherently forming a traveling band \cite{chate_2008}.
Motion of micro-organisms in a supposed to be viscous medium can be modeled using overdamped active Brownian particles (ABPs) \cite{cates1,roman,mishra2,fily_12,redner_13,digre_18,paul2022}. Such a system of ABPs, even with a repulsive interaction and in absence of any alignment interaction among them shows clustering when the density is higher than a threshold value.  This is formally known as motility induced phase separation (MIPS) \cite{cates1,mishra2,fily_12,redner_13}, resembling a vapor-liquid-like phase transition \cite{majumder_2011} or clustering in  granular gas \cite{paul_2014}. However, due to the stochastic nature of the propulsion direction, for spherical ABPs or similar non-polar particles, any kind of orientational or velocity ordering during MIPS is unlikely. For non-spherical particles, e.g., dumbells, rods, elongated micro swimmers, where the microscopic isotropy is broken, one observes velocity ordering even in absence of any explicit alignment interaction among them \cite{aranson2003,peruani2006,ginelli2010,deseigne2010,dauchot2016,bhatta2019}. Velocity ordering has been observed for spherical particles in presence of explicit alignment interaction and with repulsive interaction \cite{lam2015self,sese2021phase}. Recently, it has also been observed even in absence of explicit alignment interaction with only repulsive interaction among the ABPs, however, in absence of any thermal noise and very high packing fraction \cite{caprini2020}.
\par
% 
% In the  literature, the alignment effect has been considered as an additional interaction among ABPs \cite{martin2018,ses_2018,spreng_20}. However, in a recent work \cite{capri_21}, for high packing fraction and  in absence of any thermal noise short-ranged orientational ordering has been shown. %Other variations of this and also of the Vicsek model is used in the literature to understand properties of many nonequilibrium active systems.
\par 
In most of the studies on ABPs in the past, the dynamics was considered to be overdamped. This is motivated by the very small sizes (between nm to $\mu$m) of real systems, viz., colloids or micro-organisms, which they are thought to be representing, and hence the inertial effect can be neglected in comparison to the drag force of the solvent. However, this approximation does not stand as a general approach. Especially, when  active particles are supposed to change velocities during their motion \cite{brown_14,scholz_18}. Interestingly, inclusion of the inertial effect changes the MIPS behavior drastically. In such a scenario, it has been shown that for sufficiently large self-propulsion force, MIPS becomes reentrant \cite{suma2014}. To observe MIPS, particles in the high density region are supposed to move slowly leading to an eventual formation of clusters. For an overdamped particle, this is possible due to slowing down of overdamped particles following a collision. However, underdamped particles can bounce back from collision without slowing down until it encounters the next collision. Hence, at large self-propulsion velocity one observes a breakdown of the MIPS. 
\par
In this context, a recent study of underdamped spherical ABP with attractive interaction observed ``flocking'' or orientational ordering without any explicit alignment interaction \cite{caprini2023}. There it has been argued that the effective alignment interaction between the particles results from the interplay between persistent
active forces and attractive interactions, which gets manifested as a flocking transition. In this Letter, we take a step ahead and show that ``flocking'' can be observed for spherical underdamped active particles, i.e., active inertial particles (AIP) in absence of any explicit alignment interaction even without attractive interactions, that is with a completely repulsive interaction only.  

% \par 
% Due to inertia the velocity of an AIP changes with time, unlike ABP. In recent years, there have been a few studies which consider active  micro-swimmers with inertia \cite{lowen_20,suma_14,mandal_19,dai_20,caprini_21,paul1_21,paul1_sm_2022}. One natural question is whether one observes MIPS with the AIPs as well? We find that even though for small inertia the MIPS with AIPs is similar to that for  ABPs, there exists a vast difference for the velocity field. Contrary to an overdamped situation in which all particles,  whether they are part of the cluster or not, remain at the same temperature, inertia leads to coexistence of  different kinetic temperatures \cite{lowen_20,mandal_19}. The particles within the cluster become ``colder'' while the non-clustered particles remain ``hotter''.  The most interesting finding of this paper is the emergence of spontaneous ordering of the velocities of active particles within the ``colder'' cluster regime even in presence of external thermal noise and without any alignment interaction. Interestingly, for larger inertia the MIPS disappears and the system remains homogeneous maintaining the same kinetic temperature throughout.
\par 
 We consider $N$ interacting spherical AIPs in space dimension $d=2$. Equations governing their translational and rotational motions are \cite{lowen_20,dai_20}
\begin{eqnarray}\label{trans_eqn}
 m_i \frac{d^2\vec{r}_i}{dt^2} &=& -\gamma_t \frac{d\vec{r}_i}{dt} -\vec{\nabla}U_i+ \vec{f}_i^{{a}} +\sqrt{2\gamma_t k_B T}\vec{\xi}_i(t),\\
	\gamma_r \frac{d\theta_i}{dt}&=&\sqrt{2\gamma_rk_BT}{\eta}_i(t).
\end{eqnarray}
 Here $m_i=m$ is the mass of the $i$-th particle, $\vec{r}_i$ represents its position and  $\theta_i$ is the propulsion direction. $\gamma_t$ and $\gamma_r$ represent the translational and rotational drag coefficients, respectively, and $T$ is the temperature. The Gaussian white noises $\vec{\xi}_i$ and $\eta_i$ have zero mean and unit variance, and are Delta-correlated over space and time. The passive interaction $U_i$ as a function of the the inter-particle distance $r$ is modeled by the repulsive Weeks-Chandler-Andersen potential \cite{wca_71}
  $V_{\rm{WCA}}(r)=4\epsilon [({\sigma}/{r})^{12}- ({\sigma}/{r})^6]+{1}/{4}$, with $r < 2^{1/6}\sigma$ and $0$ otherwise. We set both diameter of the particles $\sigma$ and the interaction strength $\epsilon$ to unity.  The active force for the $i$-th particle is defined by $\vec{f}_i^a=f^a \hat{n}_i$, where $f^a$ and $\hat{n}_i \equiv (\cos \theta_i, \sin \theta_i)$ represent, respectively, the strength and the direction of activity or  self propulsion.  The translational and rotational diffusion coefficients are given as $D_{t,r}={k_BT}/{\gamma_{t,r}}$. Here, we keep  the  ratio $D_t/D_r=\sigma^2/3$ fixed with   $\gamma_t=5.0$. In this context, there are two time scales, viz., the persistence time $\tau_p=1/D_r$ defining the characteristic time after which the propulsion direction of a particle alters and the inertial time $\tau_m=m/\gamma_t$ representing the time required for a particle to achieve its terminal speed. As a measure of inertia of the particles, we define a parameter $\kappa={\tau_m}/{\tau_p}~(=m D_r/\gamma_t)$, which is varied by changing $m$. 
% Another important time scale is the mean free time of collision $\tau_c=\pi \sigma/(4f^a\phi)$ for the particles, where $\phi=N\pi \sigma^2/(4L^2)$ is the packing fraction. 
Unless otherwise mentioned, we simulate a system with a packing fraction $\phi=N\pi \sigma^2/(4L^2)=0.39$ at $T=0.1$ in a square box of linear dimension $L=128$ with periodic boundary condition in both directions. We apply the velocity-Verlet algorithm to integrate the equation of motions \cite{frenkel}, where the time step is chosen to be $0.01$ in units of $\tau_0=\gamma_t \sigma^2/\epsilon$. In the rest of the paper activity is expressed in terms of the dimensionless P\'eclet number $Pe= f^a\sigma/k_BT$, the ratio between the active energy $f^a\sigma$ and thermal energy $k_BT$ \cite{lowen_20,elgeti_15}. Unless otherwise mentioned, the subsequent results presented are for  $Pe=150$. 

% The initial positions and velocities of all the particles are chosen randomly. The initial orientations are also random with $\theta_i \in (0,2\pi)$. \\

\par
\begin{figure}[t!]
	\centering
	\includegraphics*[width=8.0 cm, height=11.0 cm]{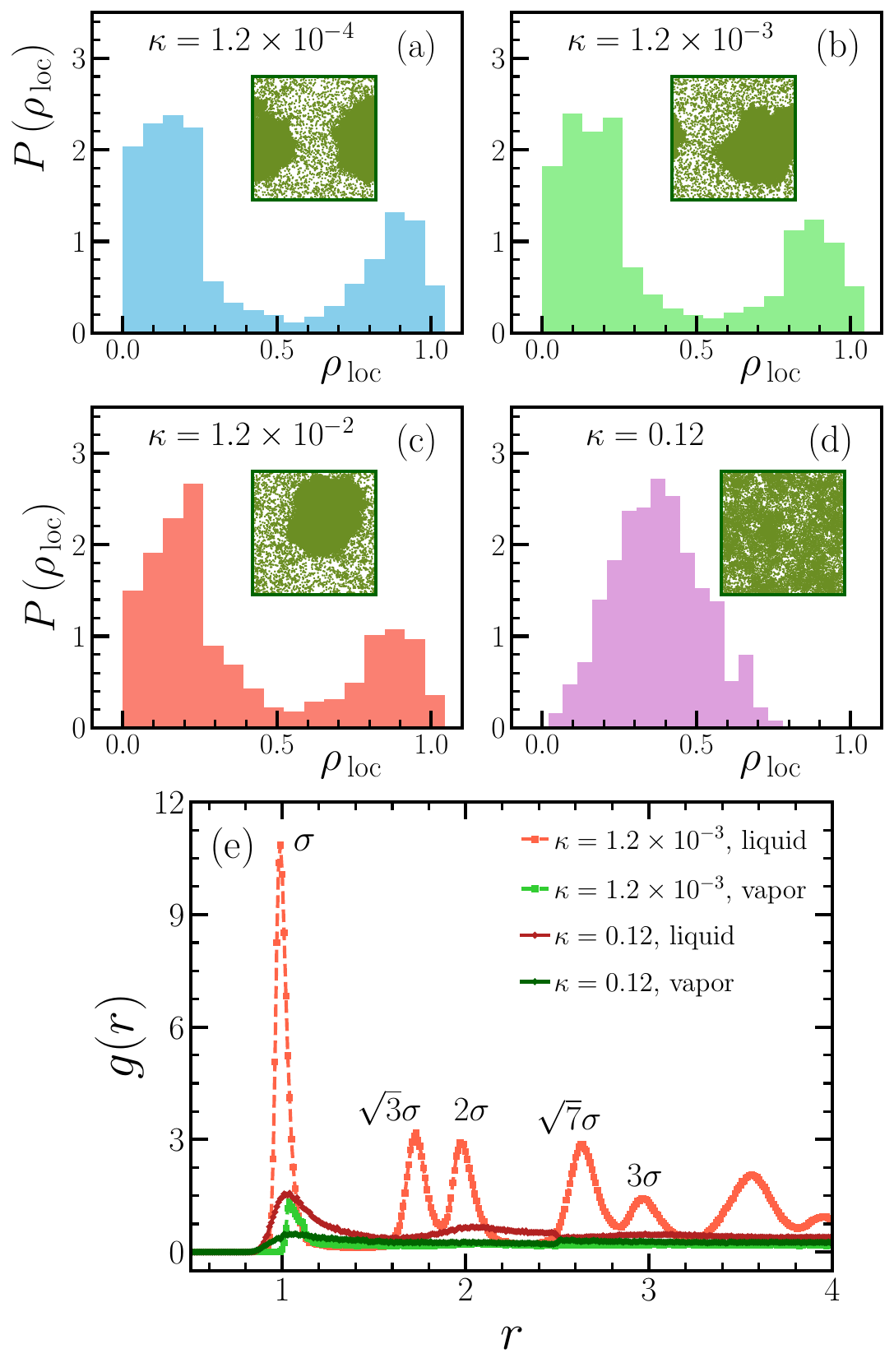}
	\caption{\label{dens_histo} (a)-(d) Normalized distributions $P(\rho_{\rm{loc}})$ of the local density $\rho_{\rm{loc}}$ for steady-state configurations at different values of the inertial parameter $\kappa$. Representatives of corresponding configurations are also shown. (e) Radial distribution function $g(r)$  considering separately the particles within the ``liquid'' and ``vapor'' phases for different values of $\kappa$. }
\end{figure}

 
 We start by showing in Fig.~\ref{dens_histo} representative snapshots of steady-state configurations which are obtained by starting from a random initial condition and letting the simulations run for a considerable amount of time until they reach a steady state. The chosen values of $\kappa$ in Figs.~\ref{dens_histo}(a)-(d) correspond to masses $m=10^{-2}, 10^{-1}, 1.0.$ and $10.0$.  The snapshots in Figs.~\ref{dens_histo}(a)-(c) clearly show phase separation between low- and high-density phases, reminiscent of the vapor-liquid phase separation in a system of passive particles with attractive interactions \cite{majumder_2011}. However, here the phase separation is solely driven by the activity in presence of a repulsive interaction among the particles, hence, basically it is an MIPS. Interestingly, in this case, unlike for ABPs, no phase separation is observed for higher values of $\kappa$, evident from the snapshot presented in Fig.\ \ref{dens_histo}(d). 
 \par
 For a quantitative understanding of the above mentioned MIPS, in Figs.~\ref{dens_histo}(a)-(d) we also show the plots of normalized distributions $P(\rho_{\rm{loc}})$ of the local densities $\rho_{\rm{loc}}$ for different values of the inertia parameter $\kappa$. To calculate $\rho_{\rm{loc}}$, at first we divide the whole simulation box into a number of smaller square boxes, each of size $4\sigma \times 4\sigma$. Then, the particle density $\rho_{\rm{loc}}$ of each of these spatially distributed boxes  is calculated. For Figs.~\ref{dens_histo}(a) and (b), i.e., for the lower values of $\kappa$, two peaks appear at $\rho_{\rm{loc}} \approx 0.15$ and $\approx 0.9$,  corresponding to the densities of the ``vapor''-like and ``liquid''-like phases, respectively. With increasing $\kappa$ the separation between the peaks gradually  decreases [Fig.~\ref{dens_histo}(c)]. For $\kappa =0.12$ the distribution becomes single peaked around $\rho_{\text{loc}} \approx 0.39$ indicating a homogeneous distribution of the particles [Fig.~\ref{dens_histo}(d)]. We have checked roughly that MIPS does not occur for $\kappa > 0.05$. Thus, for the rest of the Letter, for MIPS we choose $\kappa$ according to this limiting value.
\begin{figure*}[t!]
	\centering
	\includegraphics*[width=16.0cm, height=9.0cm]{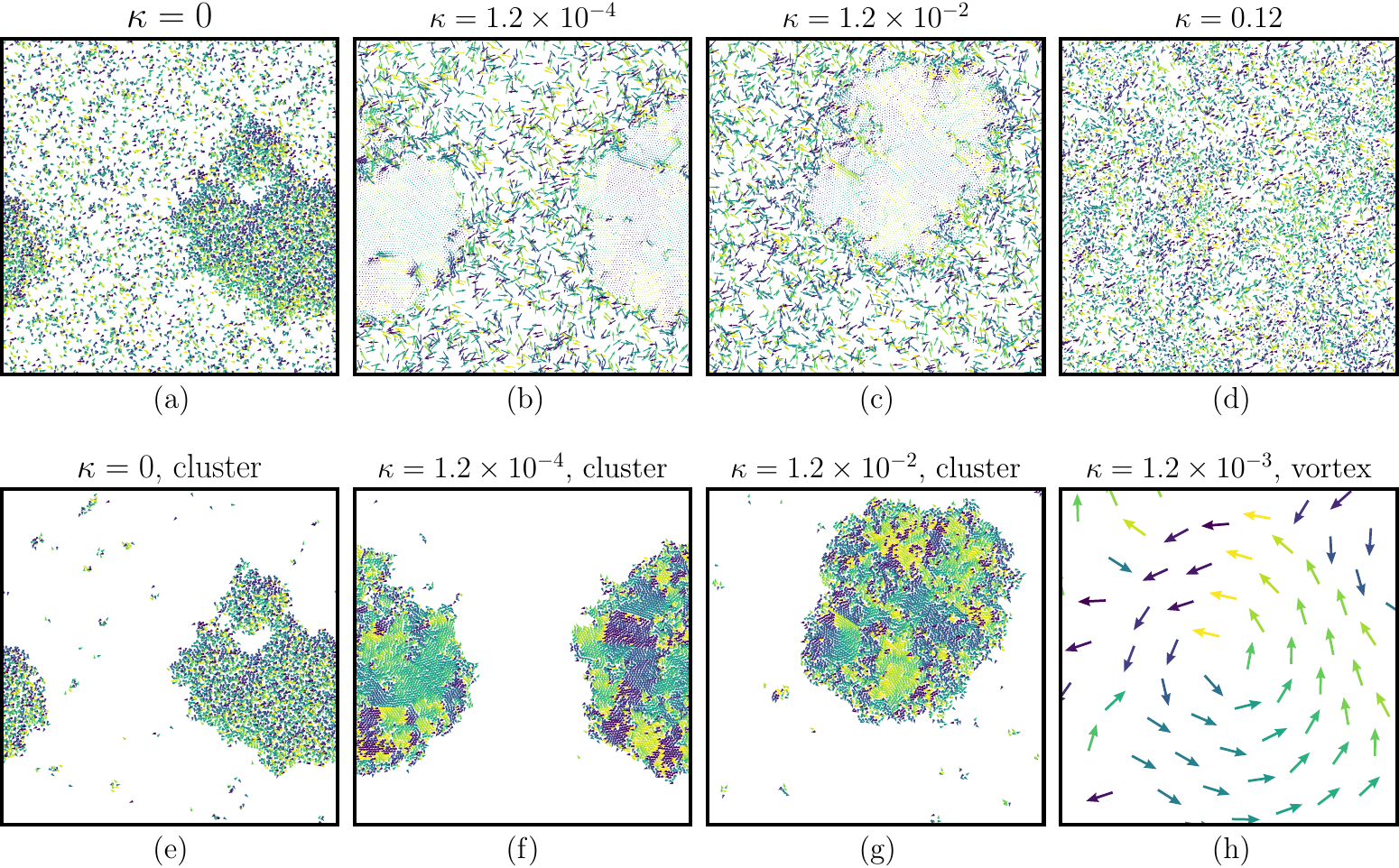}
	\caption{\label{velo_vec} (a)-(d) Typical velocity fields of all the particles for different values of $\kappa$. (e)-(g) Velocities of the particles only within the clustered region for $\kappa=0$, $1.2 \times 10^{-4}$ and $1.2 \times 10^{-2}$. (h) Section of a cluster showing a typical vortex-like defect for   $\kappa=1.2\times10^{-3}$.}
\end{figure*}
\begin{figure}[b!]
	\centering
	\includegraphics*[width=8.0cm, height=6.0cm]{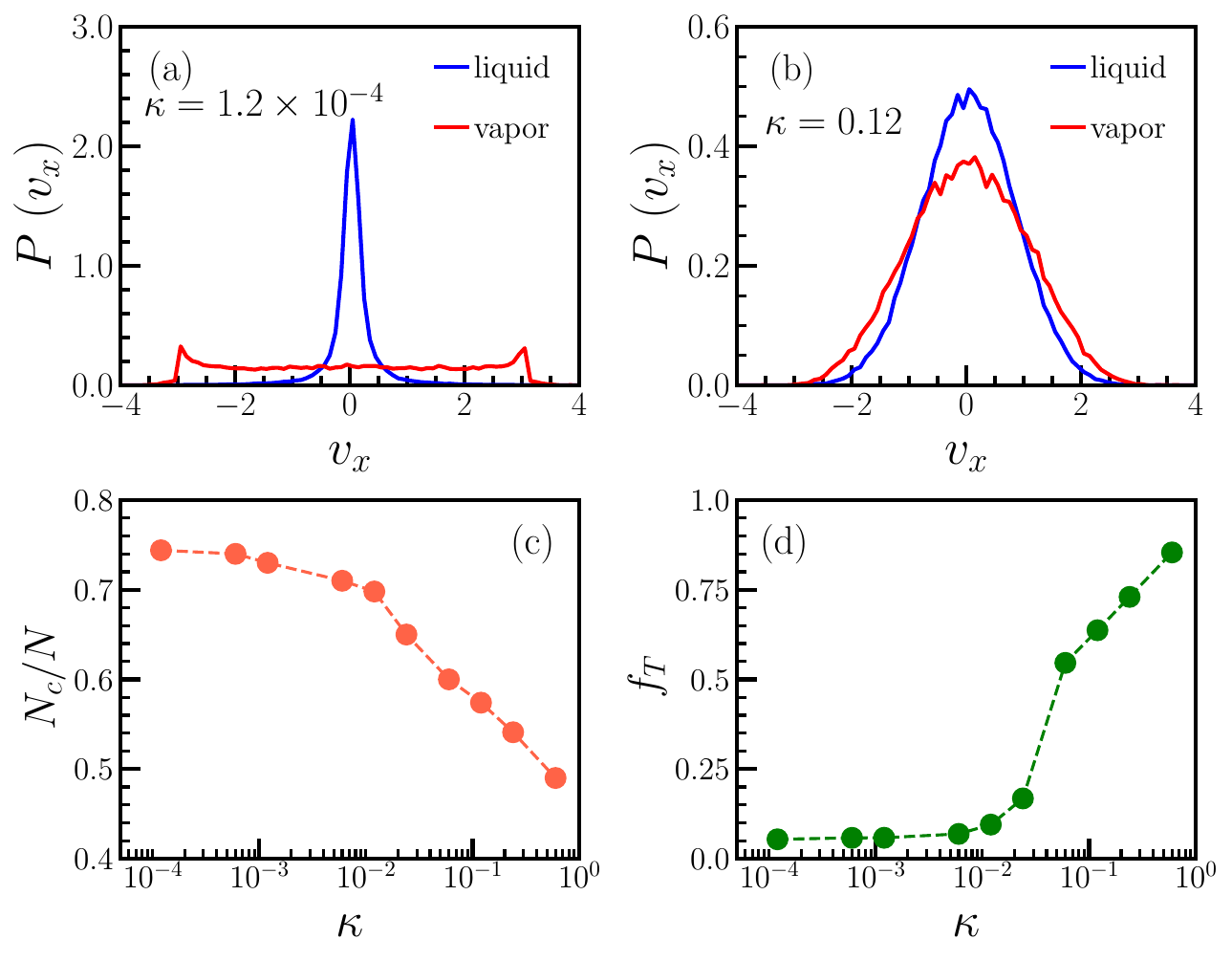}
	\caption{\label{velo_dist} (a)-(b) Distributions of the $x$-component of velocity $P(v_x)$ of particles for high and low density regions, marked as ``liquid'' and ``vapor'', respectively, for two different $\kappa$. (c) Fraction of particles that are part of any cluster as a function of $\kappa$. (d) Ratio $f_T$ between kinetic temperatures of the ``liquid'' and ``vapor'' regions versus $\kappa$. Error bars in (c) and (d) are of the size of the data points.}
\end{figure}
\par
To have an understanding of the structure formation of the observed MIPS, we calculate the radial distribution function 
$g(r)={n(r)}/{(2\pi r \delta r)}$,
where $n(r)$ counts the number of particles within a shell of width $\delta r$. In Fig.~\ref{dens_histo}(e) we show plots of $g(r)$ separately for ``vapor'' and ``liquid'' phases for two different values of $\kappa$. For $\kappa=1.2\times10^{-3}$, in the liquid-like phase, one sees different peaks, well separated from each other.  The peaks appearing at $1,~\sqrt{3},~2,~\sqrt{7},~3, \dots$ in units of $\sigma$ suggest a hexagonal quasi-periodic arrangement of the particles. Also, finite values of $g(r)$ for $r < \sigma$ indicate inter-particle separations smaller than $\sigma$ which in turn implies the existence of an effective attraction between the particles in the liquid-like phase. Since only a repulsive force is acting between the particles, it can be inferred that the effective attraction arises owing to the activity or self propulsion \cite{das2020}. On the other hand, for the ``vapor'' phase we see only one peak with much lower height followed by a constant value of $g(r)$ close to $0$. This indicates random arrangements of the particles in the ``vapor'' region.  We also plot $g(r)$ for $\kappa=0.12$ where no MIPS is observed. There, the ``liquid'' and ``vapor'' correspond to particles having a local density $\rho_{\rm{loc}}$ higher and lower than the threshold value $\rho_c=0.5$, respectively. Since the system is almsot homogeneous, for both of them  we observe only a single peak near $r \approx \sigma$ followed by a flat region, and the difference in the peak heights is insignificant compared to the case of  $\kappa=1.2\times 10^{-3}$.
\begin{figure*}[t!]
	\centering
	\includegraphics*[width=17.50cm, height=5.0cm]{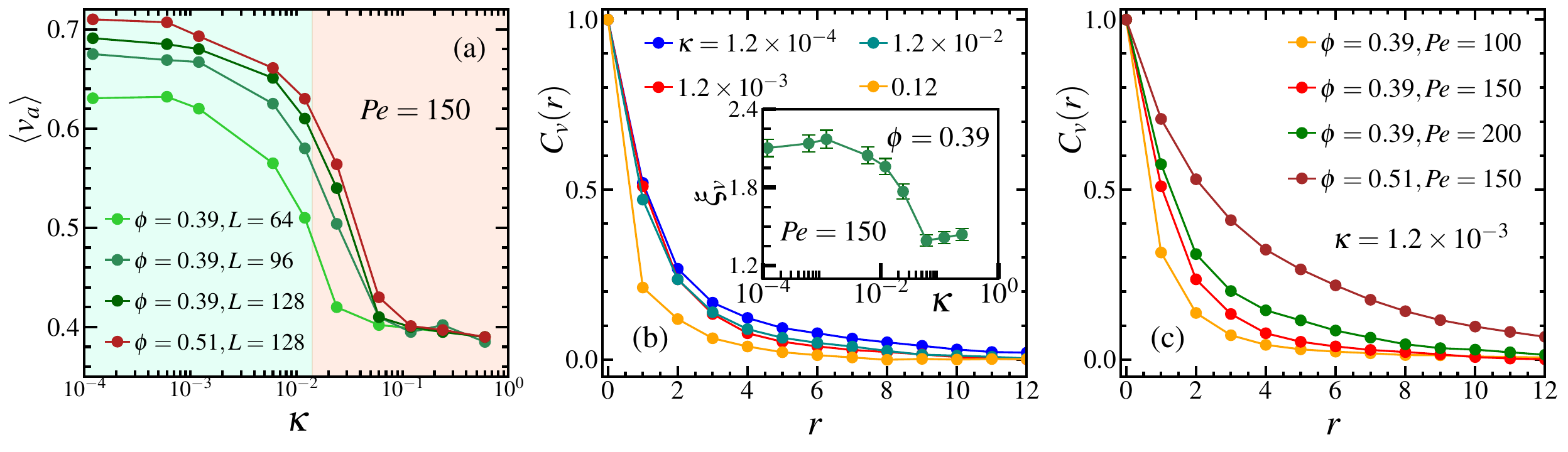}
	\caption{\label{velo_crl} (a) Velocity order parameter $\langle v_a \rangle$ versus the inertia parameter $\kappa$ for different choices of the packing fraction $\phi$ and system size $L$. Different colored regions are guide to eye. (b) Normalized velocity correlation $C_v(r)$ for different values of $\kappa$ for $\phi=0.39$ and $Pe=150$. Inset shows the variation of the correlation length $\xi_v = \int C_v(r) dr$ vs $\kappa$. (c) shows $C_v(r)$ for different choices of $\phi$ and the activity parameter $Pe$ for a fixed $\kappa=1.2\times10^{-3}$.}
\end{figure*}

\par 
 Moving on to our primary interest, in Figs.~\ref{velo_vec}(a)-(d) we show the velocity field of the particles for increasing $\kappa$. For $\kappa=0$ we consider the components of the vectors provided by $\theta_i$ as ($f^a\cos\theta_i, f^a\sin\theta_i$) whereas for $\kappa > 0$ we show the velocity components $(v_i^x,v_i^y)$. For $\kappa=0$, the particles within or outside the cluster move with same magnitude of velocity $f^a$. Now for $\kappa >0$, but with lower values, as observed from Figs.~\ref{velo_vec}(b) and (c), interestingly, the particles within the cluster have much lower velocities compared to the ones outside. For even a higher value of $\kappa=0.12$, as shown in Fig.~\ref{velo_vec}(d), where there is no MIPS the velocities of all the particles are also uniform.  In Figs.~\ref{velo_vec}(e)-(g) we show the velocities of only the clustered particles, for $\kappa=0$, $1.2\times10^{-4}$ and $1.2\times10^{-2}$, respectively. There, for a better visualization, we fixed the magnitude of the velocity vector of each particle to unity. Different colors represent distinct directional angles ranging from $-\pi$ to $\pi$. With these one clearly observes the coexistence of different domains where the particle velocities are ordered. The number of domains is larger and the fluctuations near the boundaries are higher for $\kappa=1.2\times10^{-2}$, suggesting lower degree of alignment for higher $\kappa$. This we will quantify later by calculating the correlation length and a suitable Vicsek-like order parameter.  Finally, in Fig.~\ref{velo_vec}(h) we show the presence of vortex-like defects at the boundaries of differently ordered domains for $\kappa=1.2\times 10^{-3}$. Thus, for a certain intermediate range of the inertial parameter $\kappa$, flocking of particles within micro domains appear. From the plots of  $g(r)$ in Fig.~\ref{dens_histo}(e) we already get a hint that the interparticle separation $r$ can be less than $\sigma$, indicating an  effective attraction even in presence of a purely repulsive interaction. This possibly helps in developing correlated velocities  among neighboring particles within the clusters.
%  It can be interesting to see whether the growth of clusters can be estimated from the annihilation of such defects with time, similar to the one observed for the kinetics of temperature quench in a $2d$-XY model. 

\par
Like the differences in the density field, there exist significant differences in  velocities as well while considering particles within and outside the cluster. This is evident from the normalized  distributions $P(v_x)$ of the $x$-component of the velocity plotted in Figs.~\ref{velo_dist}(a) and (b), respectively, for $\kappa=1.2\times10^{-4}$ and $0.12$. Like in Fig.~\ref{dens_histo}(e), here also we present two different data sets corresponding to the distributions of the particles in ``liquid'' and ``vapor'' regions, i.e., for clustered and non-clustered regions.  For $\kappa=1.2\times10^{-4}$,  the non-clustered ``vapor'' particles show almost a 
flat distribution, while the one for the clustered particles appears to be single-peaked near $v_x \approx 0$. However, for $\kappa=0.12$ as there is no MIPS, the distributions $P(v_x)$ for both the so-called ``vapor'' and ``liquid'' seem to be normal distributions, expected for a homogeneous system. To check  what fraction of the total number of particles becomes part of the ``liquid'' region or the cluster, for different values of $\kappa$, in Fig.~\ref{velo_dist}(c) we plot the fraction $N_c/N$ versus $\kappa$, where $N_c$ is the number of particles that are part of any cluster. We see that for  $\kappa < 1.2 \times 10^{-2}$ this number is $\approx 0.75$. Then with increasing $\kappa$, $N_c/N$ decreases and for $\kappa > 6.0\times10^{-2}$ this fraction tends towards $ \approx 0.45$ when there is no MIPS. 
\par
Even though we find only micro domains of ordered velocities, the distributions in Fig.\ \ref{velo_dist}(a) prompted us to calculate the effective temperatures defined  from the corresponding average kinetic energy as
\begin{equation}
	T_{\rm{eff}}^{l,v}=\frac{m}{2k_B}\Big[\langle v_x^2\rangle+ \langle v_y^2\rangle\Big]_{l,v}\,,
\end{equation}
where $T_{\rm{eff}}^{l}$ and $T_{\rm{eff}}^{v}$ are, respectively, the effective temperatures of the ``liquid'' and ``vapor'' regions, and $\langle \dots \rangle$ represents average over particles in those regions. In Fig.~\ref{velo_dist}(d) we plot the ratio $f_T=T_{\rm{eff}}^l/T_{\rm{eff}}^v$ as function of $\kappa$. We see that $f_T$ increases monotonically from $\approx 0.5$ toward $1$ with increasing $\kappa$. While a small value of $f_T$ indicates signficant difference in the effective temperatures of the ``liquid''-like and ``vapor''-like regions, $f_T$ approaching unity indicates that the difference becomes negligible. $f_T$ shows a jump around $\kappa \approx 0.05$ which nicely coincides with the value above which MIPS does not appear.
\par
To understand the local velocity ordering we define a Vicsek-like order parameter, related to the orientation of the particles as \cite{vicsek_95,paul1_sm_2022}
\begin{equation}\label{op}
   v_a  = \frac{1}{N_c} \sum_{i=1}^{N_c} \frac{\Big|\sum_{k=1}^{n_i}\vec{v}_k^{~i}\Big|}{\sum_{k=1}^{n_i}\big|\vec{v}_k^{~i}\big|}. 
\end{equation}
In Eq.\ \eqref{op} the summation runs over all $N_c$ particles which are part of a cluster. For each of this particle $i$, $n_i$ denotes the number of particles that are within a distance $r_c=2.5\sigma$ from it. Even though the velocities are not globally aligned in the clustered region, the order parameter $v_a$ is sufficient to capture the formation of micro domains of ordered velocities. This is demonstrated in Fig.\ \ref{velo_crl}(a) where we present the variation of $\langle v_a \rangle$ as function of the inertial parameter $\kappa$ for different packing fractions $\phi$ and lattice sizes $L$. There $\langle \dots \rangle$ denotes a steady-state average over different micro clusters and initial realizations. It clearly reflects a micro-flocking transition which becomes more prominent as $L$ increases. Of course, an increase in the particle density also sharpens the transition as shown by the data for $\phi=0.51$ for  $L=128$. The different colored regions are somewhat a guide to the eye as the system moves from a velocity ordered state to a state with random velocity ordering with increasing $\kappa$.
% \begin{figure}[t!]
% 	\centering
% 	\includegraphics*[width=8.0cm, height=7.5cm]{fig7.eps}
% 	\caption{\label{msd_part} (a)-(d) Plots of MSD of a single particle for different values of the inertia parameter $\kappa$. In each of the plots, different data sets correspond to the particles for which $\phi_{\rm{loc}}$ is lower or higher than the cut-off $\phi_c$. For all of them, the dashed and solid lines represent the power-law with the corresponding exponents $2$ and $1$.}
% \end{figure}

\par 
To further substantiate the micro-flocking transition in terms of a length scale, we calculate the two-point velocity-velocity correlation function
\begin{equation}
C_v(r)=\frac{\langle \vec{v}_i \cdot \vec{v}_j \rangle -\langle \vec{v}_i \rangle \cdot \langle \vec{v}_j \rangle}{\langle v_i^2\rangle -\langle v_i\rangle ^2} \,,
\end{equation}
where $r=|\vec{r}_i-\vec{r}_j|$ is the scalar distance between particles $i$ and $j$. In Fig.~\ref{velo_crl}(b) we plot  $C_v(r)$ for different values of $\kappa$ in the steady states. For smaller $\kappa$ where MIPS exists, we consider particles that are part of the ``liquid'' phase, whereas for larger $\kappa$ all the particles of the system are taken into account. Faster decay of $C_v(r)$ with increasing $\kappa$ clearly indicates the presence of a growing length scale capturing the micro-flocking transition. As a quantitative estimation we measure the correlation length $\xi_v =\int C_v(r) dr$ 
%\begin{equation}\label{correl_leng}
%	\xi_v =\int C_v(r) dr\,,
%\end{equation}
and show its variation as a function of $\kappa$ in the inset of Fig.\ \ref{velo_crl}(b). The behavior is very similar to what is observed for $\langle v_a \rangle$ in Fig.\ \ref{velo_crl}(a), indicating the micro-flocking transition as a function of $\kappa$. \par
To understand the effect  of packing fraction and activity strength on the velocity ordering,  in Fig.~\ref{velo_crl}(c) we plot $C_v(r)$  for different combinations of $Pe$ and $\phi$ for $\kappa=1.2\times10^{-3}$. The slower decay of $C_v(r)$ with increasing activity as well as with packing fraction indicates higher range of velocity ordering of the particles within the micro domains. A detailed quantification of $\langle v_a \rangle$ and $\xi_v$ for a complete  understanding of the observed micro flocking with variation of $\phi$ and $Pe$ will be presented elsewhere.

%\par 
%To check how the inertia parameter $\kappa$, i.e., the ratio of two timescales $\tau_m$ and $\tau_p$, determines the clustering as well as the velocity orderings. As $\kappa$ is a coupling term between mass $m$ and the rotational diffusion $D_r$. For $\kappa=0.01$, keeping $m$ fixed at $10$ we reduce the value of $D_r$ to $...$. 

\par 
In conclusion, we have reported the first observation of micro-flocking behavior in a system of purely repulsive AIPs in presence of thermal noise but absence of any explicit  alignment interaction. Previously, in Ref.~\cite{caprini2023} in presence of attractive interactions a flocking transition with AIPs has been observed, which was conjectured to be an outcome of the interplay of the attractive force with the strong persistence motion at high activities. Thus, our results challenge the generality of this conjecture and rather indicate that flocking is observed due to an effective attraction among the AIPs solely caused by the presence of strong self-propulsion forces \cite{das2020}.

Our findings are relevant for any macroscopic living objects  where inertial effects cannot be ignored. For small values of $\kappa$, the observed MIPS (for the density phase separation) is similar to what is expected for overdamped ABPs. However, interestingly, within the ``liquid'' phase, there appear vast differences in  the velocity field with  emegence of micro domains within which the velocities of AIPs are locally ordered. Also, the effective temperatures of the ``liquid'' and ``vapor'' regions appear to be different, unlike for an ABP system. For much larger $\kappa$, the MIPS behavior is no more observed, and thereby the micro-flocking transition also disappears. 
We have shown that the variations of the velocity order parameter and correlation length, with respect to $\kappa$, can be used as quantitative probes for this micro-flocking transition.
Overall, our results eastablish the fact that micro-flocking of particles is rather a generic phenomenon. Thus it will be interesting to look at the interplay between the effects of $Pe$ and $\kappa$ on such velocity ordering. As a future work, we plan to study the steady-state phase diagram in this regard. 
% \par 

\par 
In future, it will also be interesting to study the nonequilibrium scaling laws \cite{majumder_20} associated with  such velocity ordering of AIPs and the role of defect annihilation. Definitely, the effect of increasing activity and packing fraction on exchanging information among the particles, as suggested in  Fig.~\ref{velo_crl}(c), is also worth to investigate. Another future endevour would be to investigate the dynamics of such a system exploring various transport properties. Considering the recent interests in the behavior of active polymers in general \cite{winkler2020,paul3_20,paul1_sm_2022,paul2022,majumder2023,elgeti_15}, it would also be intriguing to study the structure and dynamics of a polymer consisting of AIPs as monomers.

\acknowledgments
S.P. thanks Debasish Chaudhuri for useful discussions and acknowledges ICTS-TIFR, DAE, Govt.\ of India for a research fellowship.  This work was funded by the Deutsche Forschungsgemeinschaft (DFG, German Research Foundation) under Grant No.\ 189\,853\,844 -- SFB/TRR 102 (Project B04) and further supported by the Deutsch-Franz\"osische Hochschule (DFH-UFA) through the Doctoral College ``$\mathbb{L}^4$'' under Grant No.\ CDFA-02-07, and the Leipzig Graduate School of Natural Sciences ``BuildMoNa''.  S.M. thanks the Science and Engineering Research Board (SERB), Govt.\ of India for funding through a Ramanujan Fellowship (file no.\ RJF/2021/000044). 

%
%\bibliography{aip_ss.bib}   

\end{document}